\documentclass[preprint,12pt,authoryear]{elsarticle}
\journal{Spatial Statistics}
\usepackage{lipsum}
\usepackage{amsmath,amsfonts,amsthm}
\usepackage{algorithmic}
\usepackage{algorithm}
\usepackage{array}
\usepackage[caption=false,font=normalsize,labelfont=sf,textfont=sf]{subfig}
\usepackage{textcomp}
\usepackage{stfloats}
\usepackage{url}
\usepackage{verbatim}
\usepackage{graphicx}
\usepackage{natbib}
\usepackage{booktabs}
\usepackage{hyperref}
\usepackage{tabularx}
\usepackage{tikz} \usetikzlibrary{shapes.geometric, arrows.meta, positioning, fit, backgrounds}
\usepackage{mwe} 

\theoremstyle{definition}

\usepackage{amssymb} 
\usetikzlibrary{positioning}
\usepackage{amsmath,amssymb}
\usepackage{tikz}
\usetikzlibrary{positioning}
\usetikzlibrary{shapes,arrows,positioning,fit,backgrounds,calc,matrix}
\usetikzlibrary{positioning, fit, backgrounds, arrows.meta}

\tikzset{
	roadmaptitle/.style={
		font=\bfseries\Large,
		align=center
	},
	dgpnode/.style={
		draw=teal!70!black,
		fill=teal!8,
		rounded corners=7pt,
		line width=0.9pt,
		align=center,
		minimum width=7.8cm,
		minimum height=1.35cm,
		font=\small
	},
	scen/.style={
		draw=teal!70!black,
		fill=white,
		rounded corners=6pt,
		line width=0.85pt,
		align=center,
		minimum width=3.25cm,
		minimum height=1.35cm,
		font=\small
	},
	hub/.style={
		draw=teal!70!black,
		fill=yellow!10,
		rounded corners=6pt,
		line width=0.85pt,
		align=center,
		minimum width=6.3cm,
		minimum height=1.25cm,
		font=\small
	},
	bayes/.style={
		draw=orange!75!black,
		fill=orange!8,
		rounded corners=6pt,
		line width=0.85pt,
		align=center,
		minimum width=3.75cm,
		minimum height=1.35cm,
		font=\small
	},
	spde/.style={
		draw=cyan!65!black,
		fill=cyan!7,
		rounded corners=6pt,
		line width=0.85pt,
		align=center,
		minimum width=3.8cm,
		minimum height=1.35cm,
		font=\small
	},
	partlabel/.style={
		font=\bfseries\small,
		text=teal!45!black
	},
	arrow/.style={
		-{Stealth[length=2.2mm]},
		line width=0.8pt,
		draw=teal!55!black
	},
	dashedarrow/.style={
		-{Stealth[length=2.2mm]},
		line width=0.8pt,
		draw=teal!55!black,
		dashed
	},
	partbox/.style={
		rounded corners=8pt,
		line width=0.65pt
	}
}

\tikzset{
	dgpnode/.style={draw, fill=cyan!10, rounded corners, align=center, minimum width=6cm, minimum height=1.5cm},
	scennode/.style={draw, fill=white, rounded corners, align=center, minimum width=2.2cm, minimum height=1.5cm},
	hubnode/.style={draw, fill=yellow!10, rounded corners, align=center, minimum width=5cm, minimum height=1cm},
	bayesnode/.style={draw, fill=orange!10, rounded corners, align=center, minimum width=2.4cm, minimum height=1.5cm},
	spdenode/.style={draw, fill=teal!10, rounded corners, align=center, minimum width=2.4cm, minimum height=1.5cm},
	myarrow/.style={-stealth, thick, draw=black!70},
	partlabel/.style={font=\bfseries\small, text=black!80},
	part1color/.style={cyan!5},
	part2color/.style={orange!10},
	part3color/.style={teal!5},
	accentline/.style={black!40}
}

\definecolor{spatial}{RGB}{230,97,0}      
\definecolor{demographic}{RGB}{153,0,153} 
\definecolor{merit}{RGB}{0,114,178}       
\definecolor{outcome}{RGB}{213,94,0}      
\definecolor{residual}{RGB}{0,158,115}    
\definecolor{graph}{RGB}{86,180,233}      
\definecolor{test}{RGB}{204,121,167}      

\tikzset{
	process/.style={rectangle, draw, fill=blue!10, text width=6em, text centered, minimum height=3em, rounded corners},
	data/.style={ellipse, draw, fill=green!10, text width=5em, text centered, minimum height=2.5em},
	spatial/.style={rectangle, draw=spatial, fill=spatial!10, text width=6em, text centered, minimum height=2.5em, thick},
	merit/.style={rectangle, draw=merit, fill=merit!10, text width=6em, text centered, minimum height=2.5em, thick},
	residual/.style={rectangle, draw=residual, fill=residual!10, text width=6em, text centered, minimum height=2.5em, thick, rounded corners},
	graph/.style={rectangle, draw=graph, fill=graph!10, text width=7em, text centered, minimum height=2.5em, thick},
	test/.style={rectangle, draw=test, fill=test!10, text width=6em, text centered, minimum height=2.5em, thick, rounded corners},
	arrow/.style={->, >=stealth, thick},
	bigarrow/.style={->, >=stealth, very thick, line width=1.5pt},
	annotation/.style={text width=8em, font=\small\itshape, align=center}
}

\usepackage{tikz}
\usepackage{xcolor}
\usetikzlibrary{shapes.geometric, arrows.meta, positioning, fit, backgrounds, calc}

\definecolor{hdr1}{RGB}{33,128,141}
\definecolor{hdr2}{RGB}{180,100,0}
\definecolor{hdr3}{RGB}{40,120,50}
\definecolor{row1a}{RGB}{232,246,248}
\definecolor{row1b}{RGB}{245,252,253}
\definecolor{row2a}{RGB}{255,248,235}
\definecolor{row2b}{RGB}{255,252,245}
\definecolor{row3a}{RGB}{235,248,236}
\definecolor{row3b}{RGB}{247,253,247}
\definecolor{accentline}{RGB}{33,128,141}

\usepackage{setspace}
\usepackage{tikz}
\usepackage{xcolor}
\usetikzlibrary{shapes.geometric, arrows.meta, positioning, fit, backgrounds, calc}

\definecolor{dgpcolor}{RGB}{180,225,230}
\definecolor{part1color}{RGB}{224,244,246}
\definecolor{part2color}{RGB}{255,243,224}
\definecolor{part3color}{RGB}{232,245,233}
\definecolor{nodecolor}{RGB}{255,255,255}
\definecolor{s2color}{RGB}{255,248,225}
\definecolor{robcolor}{RGB}{243,240,251}
\definecolor{accentline}{RGB}{33,128,141}
\definecolor{headtext}{RGB}{25,101,112}

\tikzset{
	dgpnode/.style={
		rectangle, rounded corners=4pt,
		fill=dgpcolor, draw=accentline, line width=1pt,
		text width=7cm, align=center,
		font=\sffamily\bfseries\small, inner sep=6pt
	},
	scennode/.style={
		rectangle, rounded corners=3pt,
		fill=nodecolor, draw=accentline, line width=0.8pt,
		text width=3.6cm, align=center,
		font=\sffamily\footnotesize, inner sep=5pt, minimum height=1.6cm
	},
	hubnode/.style={
		rectangle, rounded corners=3pt,
		fill=s2color, draw=accentline, line width=1pt,
		text width=5.5cm, align=center,
		font=\sffamily\footnotesize, inner sep=6pt, minimum height=1.4cm
	},
	bayesnode/.style={
		rectangle, rounded corners=3pt,
		fill=part2color, draw=accentline!70, line width=0.8pt,
		text width=3.4cm, align=center,
		font=\sffamily\footnotesize, inner sep=5pt, minimum height=1.5cm
	},
	spdenode/.style={
		rectangle, rounded corners=3pt,
		fill=part3color, draw=accentline!70, line width=0.8pt,
		text width=3.2cm, align=center,
		font=\sffamily\footnotesize, inner sep=5pt, minimum height=1.5cm
	},
	partlabel/.style={
		font=\sffamily\bfseries\small, text=headtext
	},
	myarrow/.style={
		-{Stealth[length=5pt]}, draw=accentline, line width=0.9pt
	}
}

\usepackage{amsmath,amssymb}
\usepackage{graphicx}
\usepackage{xcolor}
\usepackage{booktabs}
\usepackage{tabularx}
\usepackage{tikz}
\usetikzlibrary{positioning,trees}

\begin{document}
	\maketitle
	\begin{frontmatter}

		\title{\textsc{B-Fair} - A Bayesian spatial individual fairness framework}
		\author[1]{Gandhi Jafta}
		\author[1]{Inger Fabris-Rotelli}
		\author[2]{Emma Ruttkamp-Bloem}
		\author[1]{Janet van Niekerk}
		\author[1]{Iketle Maharela}

		\affiliation[1]{organization={Department of Statistics, University of  Pretoria},
			country={South Africa}}
		
		\affiliation[2]{organization={Department of Philosphy, University of Pretoria},
			country={South Africa}}
\begin{abstract}
	Spatial dependence  is often ignored in algorithmic fairness, even though location may be tied to both protected attributes and merit-relevant features. This creates a basic problem for standard fairness metrics, because peer comparisons can be distorted by spatial structure rather than by the substantive relationship we want to study. To address this, we introduce Bayesian Individual Spatial Fairness (\textsc{B-Fair}), a framework that defines fairness at the individual level while explicitly modeling protected context as a continuous stochastic spatial surface. In \textsc{B-Fair}, the protected demographic surface is treated as a latent spatial process, and the resulting uncertainty is carried through to the final fairness estimate. The \textsc{B-Fair} functional couples peer similarity, demographic disparity, and outcome differences into a single posterior quantity, so that fairness is not reduced to a point estimate but expressed as a distribution with uncertainty. Since neither the protected-context model nor the peer graph depends on administrative zones, the framework is less sensitive to areal aggregation and avoids the usual zone-based critique. We illustrate the method using the ProPublica recidivism dataset and show how our method reaches similar conclusions to those of ProPublica but with a continuous, uncertainty aware individually grounded fairness audit.
		\end{abstract}
		
		\newtheorem{proposition}{Proposition}
		
	\end{frontmatter}

	\section{Introduction}
	Algorithmic decision-making increasingly shapes spatially structured environments, including credit allocation, policing, and healthcare access. This in turn leads to a strong geographic dependency of protected attributes such as race and gender and the eventual outcome. Meanwhile, standard fairness audits typically either (i) aggregate decisions over predefined spatial units to apply group-disparity metrics or (ii) treat individuals as exchangeable, ignoring demographic context.  Both strategies are fragile in spatially stratified populations. Moreover, zone-based metrics suffer from the Modifiable Areal Unit Problem (MAUP) \citep{openshaw1983modifiable}, meaning that fairness scores fluctuate with arbitrary boundary choices (the zoning effect) and geographic scale (the scaling effect) \citep{fotheringham1991modifiable}. Furthermore, standard point-estimate metrics ignore inferential uncertainty, which is often acute for small demographic cells and minority groups. Existing spatial fairness frameworks address only some of these challenges. \par
 Continuous spatial formulations, when they exist, typically treat the protected attribute as a fixed spatial covariate rather than a latent stochastic field \citep{saxena2025legally,sacharidis2023auditing}. When space is treated as a fixed covariate there is no mechanism to fully capture the  spatial dependence in the spatial domain \cite{louzada2021spatial}. When modelled as a stochastic process and not a fixed covariate, the spatial random field is able to account for  the spatial correlation or heterogeneity over a region. Further with a random field, there exists a correlation matrix that quantifies the dependencies between random variables at different spatial locations and determines the full covariance structure of the stochastic process. Common individual fairness metrics, such as fairness through awareness and its Lipschitz-based extensions \citep{dwork2012fairness,zemel2013learning,pfaff2021fast} and counterfactual individual fairness \citep{kusner2017counterfactual}, focus on pairwise consistency between individuals but do not account for the geographic processes that shape how the data is generated. Moreover, fairness metrics are based on a point estimates, this limits the metrics ability to quantify uncertainty about decisions being made in high consequence environments. To not quantify the level of certainty when deciding if a decision is fair or not, lacks epistemic humility and could cause irreparable reputational harm to those on the receiving end of the unquantified decision. With equal concern is if a model is deemed `fair', the uncertainty still needs to be quantified in an audit.  \par
	To bridge these gaps, we introduce \emph{Bayesian Spatial Individual Fairness} (\textsc{B-Fair}) a novel continuous-space individual fairness framework based on Bayesian modelling. \textsc{B-Fair} makes the following main contributions to the literature.
	\paragraph{Continuous fairness functional for spatial data}
	\textsc{B-Fair} defines individual fairness as a continuous functional over a spatially indexed population. It operates on a fuzzy weighted peer graph induced by a task-specific distance in a residualised feature space instead of discrete groups or fixed similarity partitions. 
	\paragraph{Spatially residualised similarity for proxy-robust comparisons}
	 To mitigate spatial proxy bias, \textsc{B-Fair}  proposes a spatial residualisation procedure, implemented via latent Gaussian Matérn models and Integrated Nested Laplace Approximations (INLA) \citep{van2023new}. The procedure removes location-driven structure from individual features prior to similarity computation. This is so that peer comparisons reflect relative merit rather than geographically elicited advantage, thereby mitigating spatial proxy bias. 
	
	\paragraph{Continuous modelling of protected attributes via spatial processes} 
	\textsc{B-Fair}, models protected attributes as a latent continuous spatial field. This is to replace discrete demographic group labels with a probabilistic demographics surface that captures spatial dependence and removes reliance on arbitrary areal units. This is a principled conceptual advance in operationalising demographic exposure by replacing the typical binary minority group membership to a Bernoulli- logit Gaussian random field formulation that captures the spatial stochastic nature of demographic composition. The binary code we advance from labels the same minority status to all individuals irrespective of where they are in their position even in an administrative zone. 
	
	\paragraph{Bayesian fairness inference under spatial dependence}
	We embed \textsc{B-Fair} in a latent Gaussian modelling framework and use INLA to obtain a posterior distribution over fairness functionals, enabling uncertainty-aware fairness auditing that avoids overconfident point-estimate metrics.

	In summary, our novel framework \textsc{B-Fair}, is the first of its kind to simultaneously define fairness at the individual level via a metric based peer structure, model the protected attributes as a stochastic  latent Gaussian field  indexed over continuous space and propagate the resulting demographic uncertainty into the final fairness conclusion. 
	
	The structure of the paper is as follows: Section \ref{sec:related works} shows the related works, in relation to the previous works in subsection \ref{subsec:positioning} we position \textsc{B-Fair} in the context of the previous works in literature. The full proposed methodology is introduced in Section \ref{sec:Method}. In Section \ref{sec:chi-application} we  illustrate the \textsc{B-Fair} using the ProPublica recidivism data \citep{propublica2017carinsurance} which investigates if minority groups pay more for insurance premiums\footnote{ProPublica which investigates cases of machine learning bias, examined the dataset to see if there was racial discrimination in auto insurance prices. \url{https://www.propublica.org/article/minority-neighborhoods-higher-car-insurance-premiums-methodology?}}. 
\section{Related works}

\label{sec:related works}

\subsection{Algorithmic fairness and spatial (in)justice}
\label{algo_fair}
Group and individual fairness are the primary paradigms in algorithmic fairness. Group fairness is used to mitigate disparate mistreatment, i.e., the unequal error or misclassification rates across different demographic groups \citep{zafar2017fairness}. Group fairness frameworks impose statistical parity conditions across protected groups, such as demographic parity, equal opportunity, or equalised odds \citep{dwork2012fairness,zafar2017fairness}. Demographic parity requires that the overall rate of positive decisions (e.g. approvals or selections) is the same across protected groups, irrespective of their actual outcomes.  While group fairness is widely used, the methods require strict discretisation of populations and often assume independence between observations \citep{hardt2016equality}, and both these requirements are problematic in a spatial setting because of the MAUP problem and the independence violation.\par
Individual fairness, in contrast, formalises the principle that `similar individuals must receive similar outcomes'
\citep{dwork2012fairness,eva2024individual}. Similarity is typically defined through a task-specific metric (like a distance measure) in feature space \citep{dwork2012fairness}. This approach shifts the focus from aggregate group statistics to pairwise consistency. Traditional individual fairness frameworks treat observations as exchangeable and spatially unstructured. The reality that individuals are embedded within geographic processes that fundamentally shape demographic contexts is rarely explicitly modelled in terms of fairness.\par
Nonetheless standard individual and group fairness provide a point estimate and do not quantify uncertainty about the fairness metric. As the first departure from single certainty-equivalent value, \citet{dimitrakakis2019bayesian} and \citet{barrainkua2024uncertainty} both argue that fairness should be evaluated under model uncertainty. \citet{dimitrakakis2019bayesian} study Bayesian versions of balance and calibration in a non-spatial setting. \citet{barrainkua2024uncertainty} present `Uncertainty matters, a multi-objective uncertainty-aware algorithmic comparison framework'. 
\subsection{Emerging frameworks in spatial fairness }
Spatial fairness research highlights that location is not a neutral attribute. Location is laden with stories of historical segregation, resource allocation, and structural inequality\citep{dadashpoor2025defining}. Machine learning models are deployed and ignore the geographic locations where they are used, the spatial variable can act as a proxy for protected features. Further, it will likely reinforce the structural inequalities that created the variable in the first place \citep{baresi2025neo,dadashpoor2025defining}.\par
Acknowledging the limitations of spatially blind models, a growing body of literature has formalised the concept of `spatial fairness'. Early approaches in this domain introduced fair polynomials and distance-based mechanisms to address specific location-based decision rules, such as distance thresholds and service zones. \citet{shaham2024fair} proposed spatial indexing algorithms to balance protected group attributes within a discrete spatial partition, with \citet{shaham2024fairkd} further formalising this approach via a Fair KD-tree construction that explicitly re-districts spatial neighbourhoods to reduce calibration error across regions. \citet{sacharidis2023auditing} introduces an auditing framework that scans the space for statistically significant discrepancies in positive outcome rates relative to neighbouring regions. \par

\citet{xie2022fairness} introduce `fairness by where' ,  model-agnostic bi-level learning framework for spatial fairness, building on the Space-As-a-Distribution training strategy of \citet{he2022sailing},which shows that model performance can be manipulated toward a wide range of fairness-bias outcomes across different space partitionings without loss of accuracy. This exposes the fragility of any single-partitioning fairness verdict. \citet{kyriakopoulos2025promis} propose a model-agnostic post-processing optimisation problem that minimises a fairness measure designed to be resilient to gerrymandering, subject to a constraint that limits the allowable deviation from the original model’s performance. Also, \citet{saxena2025legally} introduced Legally Compliant (LC) Spatial Fairness, which helps distinguish between outcomes that depend on where people live and those that specifically affect legally protected groups. Most recently, \citet{saxena2024spatial} independently arrive at the same critique in a position paper surveying this literature, arguing that existing spatial fairness work rarely verifies bias mitigation against the true underlying protected attribute and treats location as a neutral, purely technical covariate rather than one embedded in historical and socioeconomic context. The reality is underlying demographic surfaces are inferred from sampled data or noisy census aggregations, making them inherently stochastic fields. Existing frameworks fail to propagate spatial estimation uncertainty into downstream fairness evaluation, risking false confidence in fairness audits where demographic density is highly uncertain.

\subsection{Positioning of this work}
\label{subsec:positioning}
This paper bridges algorithmic fairness and spatial statistics by introducing a continuous-space individual fairness framework that explicitly models demographic context as a latent spatially varying process, represented via a Matérn  random field approximated through the SPDE approach \citep{lindgren2011explicit}. Different to group-based disparity measures, \textsc{B-Fair} does not rely on administrative partitions or aggregate counts over predefined zones. Unlike individual fairness, it embeds similarity-based peer comparisons within a structured spatial field rather than treating individuals as exchangeable. Unlike purely spatial clustering or hotspot detection methods, \textsc{B-Fair} defines a normative fairness functional and equips it with a probabilistic inferential apparatus to quantify uncertainty about fairness violations.\par
By combining feature-space peer comparison with latent Gaussian spatial modelling \citep{lindgren2011explicit} and point-estimate and posterior-based violation metrics, \textsc{B-Fair} integrates methodological developments from fairness theory, spatial statistics, and Bayesian inference into a unified framework for continuous-space fairness auditing.
	\section{Methodology}
	\label{sec:Method}
	In this section we begin by introducing the \textsc{B-Fair} notation in Section \ref{sec:setup}.  This is followed by the an in-depth explanation of how merit based features are used to create a fuzzy peer graph in Section \ref{sec: meritbased}. Section \ref{sec:protected} shows how we construct the protected attribute model. The \textsc{B-Fair} functional which combines the merit based peer structure, protected attribute model and the outcome difference is then defined in Section \ref{sec: algorithmic disparity}. Lastly, we present the posterior fairness distribution in Section \ref{sec: post fairness}. 
	
	\subsection{Notation and setup}
	\label{sec:setup}
	Let $\mathcal{S} \subset \mathbb{R}^2$ denote the spatial domain where $n$ independent individuals are observed. Each individual $i$ is characterised by:
	\begin{itemize}
		\item $s_i \in \mathcal{S} $ - geographic location 
		\item $\mathbf{f}_i = (f_i^{(1)}, \ldots, f_i^{(k)}) \in \mathbb{R}^k$ - raw merit feature
		vector for person $i$ where $k$ is the total number of features
		\item $O_i \in [0, 1]$ -  algorithmic outcome probability, or else  $O_i \in \{0, 1\}$  if only binary outcome available
		\item $A_i \in \{0, 1\}$ - observed protected attribute
		(e.g.\ $A_i = 1$ if the individual is a minority group member)
	\end{itemize}
	The complete dataset is $\mathcal{D}_n = \{(\mathbf{f}_i, O_i, s_i, A_i)\}_{i=1}^{n}$. We write $\mathbf{f_i} \in \mathbb{R}^{n \times k}$ for the raw feature matrix and $\mathbf{O} = (O_1, \ldots, O_n)$ for the model outcome vector.
	The \textsc{B-Fair} framework can be summarised in three main components: 
	\begin{itemize}
		\item \textbf{Account for spatial correlation when constructing peers} 
		We residualise merit features via latent Gaussian Matérn models \citep{lindgren2011explicit,dupont2022spatialplus}, isolating task-relevant structure from spatial dependence, and use the resulting residualised feature space to build a fuzzy peer graph via Mahalanobis similarity geometry.
		\item \textbf{Spatial model for the protected attributes} 
		We represent the protected attribute as a latent Gaussian random field (GRF), yielding a probabilistic protected-context surface that captures spatial dependence and demographic uncertainty.
		\item \textbf{Evaluate and quantify stochastic fairness} 
		We define a fairness functional that weighs outcome discrepancies between peers by their demographic contrasts and propagates joint posterior draws of the merit and demographic fields through it, thereby producing a posterior distribution of the fairness metric.
	\end{itemize}

	\subsection{Merit based feature and peer construction}
	\label{sec: meritbased}
	We formalise \textsc{B-Fair}  by evaluating fairness relationally, by comparing individuals within a metric feature space defined over merit-based (non-protected) attributes. Here, merit-based attributes are the non-protected features relevant to the decision on an individual level. To define this feature space, we draw on the individual fairness principle of \citet{dwork2012fairness}, which states that similar individuals should be treated similarly. This principle assumes the existence of a population-level similarity metric 
	$d : \mathcal{F} \times \mathcal{F} \rightarrow \mathbb{R}_{+}$ over the merit-based feature space $\mathcal{F} \subseteq \mathbb{R}^k$, where smaller values of $d(\mathbf{f}_i, \mathbf{f}_j)$ indicate greater similarity 
	between individual feature vectors $\mathbf{f}_i, \mathbf{f}_j \in \mathcal{F}$ 
	\citep{dwork2012fairness, eva2024individual}. \par

	Directly applying a similarity metric to individual feature vectors 
	$\mathbf{f}_i \in \mathbb{R}^k$  as observed risks spatial confounding. Spatial confounding arises when the covariates of interest, in our case the merit features, are spatially dependent but not fully determined by their spatial location alone \citep{dupont2022spatial}.  Many socioeconomic variables, such as income, housing characteristics, or access to services, implicitly encode geographic location, so that the combination of feature values and spatial coordinates can act as a proxy for protected attributes. Comparing individuals in this raw feature space therefore conflates genuine merit similarity with shared spatial and demographic context, obscuring whether outcome differences are attributable to merit or to protected spatial structure. To mitigate this, we remove the spatially structured component of the merit features via spatial residualisation before computing pairwise similarity, so that peer comparisons reflect merit net of location. \par
	We model each raw merit feature using a latent GRF with a Matérn covariance function to represent the spatial structure and its uncertainty explicitly \citep{opitz2017latent}. For each individual $i$ at geographic location $s_i$, the raw merit feature $\mathbf{f_i}$ is modelled as

	\begin{equation}
		\mathbf{f_i} = \mu + \mathbf{\xi}(s_i)+ \varepsilon_i \quad \varepsilon_i\sim \mathcal{N}(0, \sigma_\varepsilon^2)
	\end{equation} 
	where $\mu$ is the global intercept for feature $p$, and $\xi(s)$ is a zero-mean spatial Matérn field. This implies that $\xi(s)$ is a continuous GRF where the a priori expected value is zero across all spatial locations $s$, and the a priori spatial correlation between any two locations decays as a function of their Euclidean distance. Lastly, $\varepsilon_i$ represents independent and identically distributed Gaussian noise. \par
	Using the INLA framework \citep{Inlaspatial2009}, we approximate the full joint posterior distribution of the spatial process and its hyperparameters, $p(\boldsymbol{\xi}, \mu, \sigma_{\varepsilon,}^2 \mid \{f_i, s_i\}_{i=1}^n)$, where $\boldsymbol{\xi} = [\xi(s_1), \dots, \xi(s_n)]^{\top} \in \mathbb{R}^n$ represents the stacked vector of latent spatial effects across all sampled locations.  Rather than relying on a single deterministic point estimate for spatial residualisation, we draw $M$ joint posterior samples from this distribution, 
	\begin{equation}
		\left\{ \left( \mu^{(m)}, \xi^{(m)}(\cdot) \right) \right\}_{m=1}^M \sim p \left(\mu, \xi, \sigma^2_\varepsilon \mid \mathcal{D}_n\right).
	\end{equation}
	For each posterior sample $m$, we compute the residualised feature  $\tilde{f}_i^{(m)}$, isolating the individual's merit-relevant attribute net of the estimated spatial surface at their specific coordinate
	\begin{equation}
		\label{eq:residualised metrit vector}
		\tilde{f}_i^{(m)} = f_i - \left[\mu^{(m)} + \xi^{(m)}(s_i)\right].
	\end{equation}
	By stacking these scalar features across all $P$ dimensions, we construct the $i'th$ residualised feature vector $\tilde{\mathbf{f}}_i^{(m)} = \left[\tilde{f}_i^{(1,m)}, \dots, \tilde{f}_i^{(m)}\right]^{\top} \in \mathbb{R}$. Sampling $\tilde{\mathbf{f}}_i^{(m)}$ generates a posterior distribution over the residualised feature space itself, allowing  uncertainty to propagate directly into the downstream similarity calculations. Using these sampled residualised merit features, we define similarity through the Mahalanobis distance \citep{mahalanobis1930tests} for each posterior draw $m$,
	\begin{equation}
		d_{\tilde{\mathcal{F}}}^{(m)}(i,j) = \sqrt{(\tilde{\mathbf{f}}_i^{(m)} - \tilde{\mathbf{f}}_j^{(m)})^{\top} (\Sigma^{(m)})^{-1} (\tilde{\mathbf{f}}_i^{(m)} - \tilde{\mathbf{f}}_j^{(m)})},
	\end{equation}
	where $\Sigma^{(m)} = \mathrm{Cov}(\tilde{\mathbf{f}}^{(m)}) \in \mathbb{R}^{k \times k}$ is the empirical covariance matrix of the residualised features across all $n$ individuals for posterior draw $m$. Under the condition that $n > k$ and that the residualised features exhibit no perfect collinearity, $\Sigma^{(m)}$ is symmetric, positive-definite and strictly invertible \cite{mardia1979principal,anderson2003introduction}. This empirical covariance matrix is estimated from the posterior samples as
	\begin{equation}
		\Sigma^{(m)} = \frac{1}{n-1} \sum_{i=1}^n \left(\tilde{\mathbf{f}}_i^{(m)} - \bar{\tilde{\mathbf{f}}}^{(m)}\right)\left(\tilde{\mathbf{f}}_i^{(m)} - \bar{\tilde{\mathbf{f}}}^{(m)}\right)^{\top},
	\end{equation}
	where $\bar{\tilde{\mathbf{f}}}^{(m)} = \frac{1}{n}\sum_{i=1}^n \tilde{\mathbf{f}}_i^{(m)}$ is the sample mean vector for draw $m$. This covariance structure reflects the joint dependence among merit features after spatial residualisation, capturing correlations that arise from socioeconomic mechanisms rather than geographic proximity. \par
	Merit features such as income, debt, and asset value typically exhibit strong collinearity, both due to underlying socioeconomic structure and the residual correlation that persists even  after spatial  residualisation \citep{Duan2025BinaryIC, de2000mahalanobis, rhee2019comparison}. The Mahalanobis distance accounts for the full joint covariance of the residualised feature space, decorrelating these dimensions and rendering the metric scale-invariant \citep{mahalanobis1930tests, mclachlan1999mahalanobis}.  Unlike Euclidean distance, which treats all feature dimensions equally and is distorted by multicollinearity \citep{de2000mahalanobis}, the Mahalanobis distance whitens the feature space by weighting each dimension by the inverse of its variance and correcting for cross-feature correlations, ensuring that similarity is measured in units of statistical distance rather than raw feature units \citep{de2000mahalanobis}. \par
	To operationalise the individual fairness principle \citep{dwork2012fairness}, we reconstruct the population's relational structure. Rather than imposing a rigid binary threshold, which would force deterministic connections between distant individuals, we define $\mathcal{P}_h$ as a fuzzy peer graph \citep{mordeson2000applications} wherein the degree of edge membership for any peer-pair $(i,j)$ is continuous. The edge weight $w_{\mathcal{P}_h}^{(m)}(i,j)$ connecting peer-pair $(i,j)$ under posterior sample $m$ is governed by a radial basis function (RBF) kernel such that the graph edge membership weight is defined by,
	\begin{equation}
		w_{\mathcal{P}_h}^{(m)}(i,j) = \exp\left(-\frac{(d_{\tilde{\mathcal{F}}}^{(m)}(i,j))^2}{h^2}\right),
	\end{equation}
	where $h > 0$ is a bandwidth parameter governing the decay rate of comparability. The bandwidth $h$ acts as a global scale parameter; in practice, we set $h$ equal to the $\tau^{\text{th}}$ percentile of the pairwise Mahalanobis distance distribution. If two individuals are mathematically identical across all residualised merit features, their graph edge membership weight is $1$. As their distance in the residualised space increases, their mutual membership in the peer graph smoothly decays toward $0$. This formulation ensures that individual fairness constraints are evaluated as a fluid relational structure over the data, mitigating the brittleness of hard categorical thresholds and preserving the full continuous gradient of peer similarity.
	\subsection{Protected attribute spatial model}
	\label{sec:protected}
	Having established a merit-based peer structure that is statistically decoupled from spatial dependence, we now model the demographic structure as a continuous spatial process. In highly segregated real-world settings, protected attributes are not distributed independently or randomly across individuals; rather, they exhibit spatial dependence driven by historical, socioeconomic, and structural factors. We model the observed protected attribute as a realisation of a latent spatial stochastic process. Recall that $s_i \in \mathcal{S} \subset \mathbb{R}^2$ denotes the exact spatial coordinate of individual $i$, and  $A_i \in \{0,1\}$ denotes their observed, discrete protected attribute (e.g., minority status). We assume that $A_i$ is conditionally distributed according to a spatially structured probability surface $\pi$. We write $ \mathbf{\pi} : \mathcal{S} \to [0,1]$ for the continuous surface and $\pi(s_i)$ for its evaluation at an individual $i'$s location. The protected surface is modelled as, 
	\begin{equation}
		A_i \mid \pi(s_i) \sim \mathrm{Bernoulli}(\pi(s_i)),
	\end{equation}
	where the linear predictor gives the log-odds of the protected demographic presence
	\begin{equation}
		\label{logit}
		\mathrm{logit}(\pi(s_i)) = \theta_0 + u_A(s_i).
	\end{equation}
	Here, $\theta_0$ acts as the global intercept (the global intercept on the log-odds scale, corresponding to the baseline log-odds of protected group membership in the absence of spatial variation), and $u_A(s)$ is a latent, continuously indexed Gaussian Random Field defined over $\mathcal{S}$. With a Matérn covariance structure the inverse-logit transformation of this process, $\pi(s)$, defines the protected attribute surface. Rather than representing a deterministic boundary, $\pi(s)$ captures the probabilistic exposure of the protected group at any arbitrary coordinate $s$. 
	The posterior distribution $P(\pi(s) | \mathcal{D}_n$) is approximated via INLA, yielding $m$ joint posterior samples of the linear predictor given in Equation \ref{logit}, which are transformed via the inverse logit to produce samples of $\pi(s_i)$ at each observed location.
	
	\subsection{Algorithmic disparity and the empirical \textsc{B-Fair} functional}
	\label{sec: algorithmic disparity}
	To quantify algorithmic divergence between any two individuals, we define the (symmetric) outcome disparity/violation function
	\begin{equation}
		V_\eta(O_i,O_j) = \max\{0,\, |O_i - O_j| - \eta\},
	\end{equation}
	where $\eta \ge 0$ is a tolerance threshold. This hinge-loss formulation downweights negligible, sub-threshold predictive variations and penalises only systematic outcome divergence. Because it depends on the absolute difference $|O_i - O_j|$, $V_\eta$ is symmetric: it detects large discrepancies in either direction and does not, by itself, distinguish which individual is disadvantaged.
	Given the protected-context surface $\pi^{(m)}(s)$ and the fuzzy merit weights  $w_{\mathcal{P}_h}$, we formalise the empirical \textsc{B-Fair} functional as a graph-weighted expectation of pairwise violations among comparable individuals.  The \textsc{B-Fair} functional admits both a symmetric and a directional variant. The symmetric version treats any large outcome discrepancy across a protected-context gradient as a violation, regardless of which side receives the worse outcome, and is well-suited as a screening statistic for the \emph{existence} of spatially structured unfairness.  The directional version incorporates a normative choice about harm direction by penalising only pairs in which the individual in a higher-protected-context region receives a worse outcome than a merit-comparable peer in a lower-protected-context region, and is more appropriate for substantive audits concerned with discrimination against high-minority contexts. In the symmetric version,  the empirical estimator of the \textsc{B-Fair} function is defined as 
	\begin{equation}
		\label{B-Fair-symmetric}
		\widehat{T}_{\mathrm{sym}}(\pi^{(m)})
		= \frac{\left(\sum_{i < j} w_{\mathcal{P}_h}^{(m)}(i,j)\, \bigl|\pi^{(m)}(s_i)-\pi^{(m)}(s_j)\bigr|\,V_\eta(O_i,O_j)\right)}
		{\sum_{i < j} w_{\mathcal{P}_h}^{(m)}(i,j)}.
	\end{equation}
	This functional captures the expected spatial fairness violation per unit of peer comparability and demographic context difference: the numerator compounds three relational dimensions-fuzzy merit similarity $w_{\mathcal{P}_h}^{(m)}(i,j)$, contextual demographic gap $\lvert \pi^{(m)}(s_i)-\pi^{(m)}(s_j)\rvert$, and algorithmic outcome penalty $V_\eta$ -while the denominator normalises by the total effective mass of the fuzzy peer graph, rendering the estimator invariant to population scaling.
	In our main analysis, we adopt the symmetric form for detection and implement directionality at the level of the \textsc{B-Fair} functional via an explicit indicator on the sign of both the context gap and the outcome difference.
	In many applications, however, the direction of the demographic gradient matters normatively. For instance, we may wish to penalise more heavily those configurations in which individuals in higher-protected-context regions (e.g., higher minority prevalence) systematically receive worse outcomes than their merit-comparable peers in lower-protected-context regions. To capture this asymmetry, we define a directional version of \textsc{B-Fair} by introducing a sign indicator
	\begin{equation}
		\label{eq:dir-indicator}
		\delta_{ij}
		=
		\mathbb{I}\bigl\{\pi^{(m)}(s_i) > \pi^{(m)}(s_j),\, O_i > O_j\bigr\}
		+
		\mathbb{I}\bigl\{\pi^{(m)}(s_j) > \pi^{(m)}(s_i),\, O_j > O_i\bigr\},
	\end{equation}
	which is $1$ precisely when the individual in the higher-protected-context setting receives the worst outcome. The directional empirical \textsc{B-Fair} functional is then defined as
	\begin{equation}
		\label{B-Fair-directional}
		\widehat{T}_{\mathrm{dir}}(\pi^{(m)})
		= \frac{\left(\sum_{i < j} w_{\mathcal{P}_h}^{(m)}(i,j) \bigl|\pi^{(m)}(s_i)-\pi^{(m)}(s_j)\bigr|\, V_\eta(O_i,O_j)\, \delta_{ij}\right)}
		{\sum_{i < j} w_{\mathcal{P}_h}^{(m)}(i,j)}.
	\end{equation}
	While $\widehat{T}_{\mathrm{sym}}$ is sensitive to any large outcome discrepancy across a protected-context gap, $\widehat{T}_{\mathrm{dir}}$ isolates those violations that align with a substantive fairness concern about worse treatment in higher-protected-context areas.
	
	Finally, we note that the contextual gap term $\bigl|\pi^{(m)}(s_i)-\pi^{(m)}(s_j)\bigr|$ introduces an explicit dependence on the scale of the protected-context surface and on the spatial smoothness of $\pi^{(m)}$. If $\pi^{(m)}$ varies  slowly over space, these gaps will be uniformly small, attenuating \textsc{B-Fair}; conversely, highly heterogeneous context surfaces can inflate violation scores. In practice, this makes the choice of spatial scale for modelling $\pi^{(m)}$ (e.g., via kernel bandwidth or Matérn range) and the normalisation of $\pi^{(m)}$ itself important modelling decisions, and \textsc{B-Fair} should be interpreted relative to these scale choices rather than as an absolute, unit-free quantity.
	\subsection{Computational considerations}
	\label{impementation note}
	Although the theoretical functionals in Equations \ref{B-Fair-symmetric}
	and \ref{B-Fair-directional} are defined over all unordered pairs
	\(i < j\), practical computation is carried out on a sparse peer graph
	for scalability.
	After residualising the feature space to remove spatial dependence, a
	directed \(k\)-nearest-neighbour (k-NN) graph is constructed using the
	Mahalanobis distance.
	Because the neighbour relation is not generally symmetric, the directed
	graph is subsequently symmetrised by taking the union of reciprocal and
	non-reciprocal neighbour links, yielding the undirected edge set
	\[
	\mathcal{E}_k
	=
	\bigl\{\{i,j\} : j \in \mathcal{N}_k(i)\ \text{or}\ i \in \mathcal{N}_k(j)\bigr\}.
	\]
	The empirical \textsc{B-Fair} functionals are then evaluated over this edge set,
	with each unordered pair \(\{i,j\}\) contributing exactly once.
	This ensures consistency with the \(i<j\) formulation in the theory and
	avoids the double counting that would arise under a purely directed
	enumeration of peers.
	The functional has an intuitive interpretation,
	\begin{itemize}
		\item If outcome differences are independent of demographic structure,
		$\widehat{T}(\pi^{(m)}) \approx 0$.
		\item If disparities increase with demographic contrast,
		$\widehat{T}(\pi^{(m)}) > 0$.
	\end{itemize}
	Thus, \textsc{B-Fair} quantifies spatially structured violations of individual fairness. It measures the degree to which algorithmically comparable individuals receive systematically different outcomes as a function of the demographic composition of their locations.\par 
	
	\subsection{Posterior fairness distribution}
	\label{sec: post fairness}
	A limitation of traditional algorithmic auditing is the reliance on
	point-estimate fairness metrics, which fail to account for the epistemic
	uncertainty inherent in observational demographic data and all the uncertainty from the model formulation. 
	\citep{dimitrakakis2019bayesian,barrainkua2024uncertainty}. Since \textsc{B-Fair} models both the protected demographic context and the merit feature
	space as latent spatial stochastic processes, we propagate uncertainty from both surfaces, which are jointly considered in the final fairness evaluation.
	Formally, let $p(\pi\mid\mathcal{D}_n)$ denote the posterior distribution of the continuous demographic surface and let
	$p(\tilde{\mathbf{f}} \mid \mathcal{D}_n)$ denote the posterior distribution
	over the residualised feature space, both conditioned on the observed data
	$\mathcal{D}_n$. Writing $T(\tilde{\mathbf{f}}, \pi)$ for either
	$\widehat{T}_{\mathrm{sym}}$ or $\widehat{T}_{\mathrm{dir}}$, the posterior  of the \textsc{B-Fair} metric is the pushforward of the joint posterior through the \textsc{B-Fair} functional,
	\begin{equation}
		p(T \mid \mathcal{D}_n)
		= T_{\#}\, p\!\left(\tilde{\mathbf{f}},\, \pi \mid \mathcal{D}_n\right),
	\end{equation}
	where $T_{\#}$ denotes the pushforward measure.
	Concretely, \textsc{B-Fair} propagates uncertainty through two independent posterior
	sources. First, for each merit feature we fit a Gaussian SPDE model
	via INLA and draw $M$ joint posterior samples
	$\tilde{\mathbf{f}}^{(m)} \sim p(\tilde{\mathbf{f}}\mid\mathcal{D}_n)$; for
	each draw the peer graph $w_{\mathcal{P}_h}^{(m)}$ is reconstructed as a
	deterministic function of $\tilde{\mathbf{f}}^{(m)}$. Second, we fit a
	binomial SPDE model for the protected attribute via INLA
	and draw $\pi^{(m)} \sim p(\pi \mid \mathcal{D}_n)$ from
	the protected-context posterior. The \textsc{B-Fair} functional is then evaluated on the
	joint draw $\bigl(\tilde{\mathbf{f}}^{(m)},\, w_{\mathcal{P}_h}^{(m)},\,
	\pi^{(m)}\bigr)$, yielding a doubly-pushforward posterior distribution
	that propagates uncertainty from both the spatial merit surface and the spatial demographic surface into the final fairness assessment. The two posteriors are
	treated as independent; a fully joint treatment would require a single
	hierarchical model over both surfaces simultaneously, which we leave for future
	work.
	
	For each posterior draw $m = 1, \dots, M$, we evaluate the empirical \textsc{B-Fair}
	functional
	\[
	T^{(m)} = T\!\left(\tilde{\mathbf{f}}^{(m)},\, \pi^{(m)}\right),
	\]
	using the draw-specific residualised features and peer graph. 
	The expectation of the spatial fairness violation is then approximated by
	\begin{equation}
		\mathbb{E}[T \mid \mathcal{D}_n]
		\approx \frac{1}{M}\sum_{m=1}^{M} T^{(m)},
	\end{equation}
	with the empirical distribution of $\{T^{(m)}\}_{m=1}^{M}$ providing credible
	intervals and posterior exceedance probabilities for global fairness statements.
	We declare evidence of spatial fairness violation when
	\[
	P\!\left(T > \tau \mid \mathcal{D}_n\right) > 1 - \alpha,
	\]
	for a pre-specified threshold $\tau$ and significance level $\alpha$, where $\tau$ is calibrated against a permutation null distribution. Figure \ref {figsec: b-fairtsc{B-Fair}-full-double} is a graphical illustration of the full double pushfoward \textsc{B-Fair} audit.

	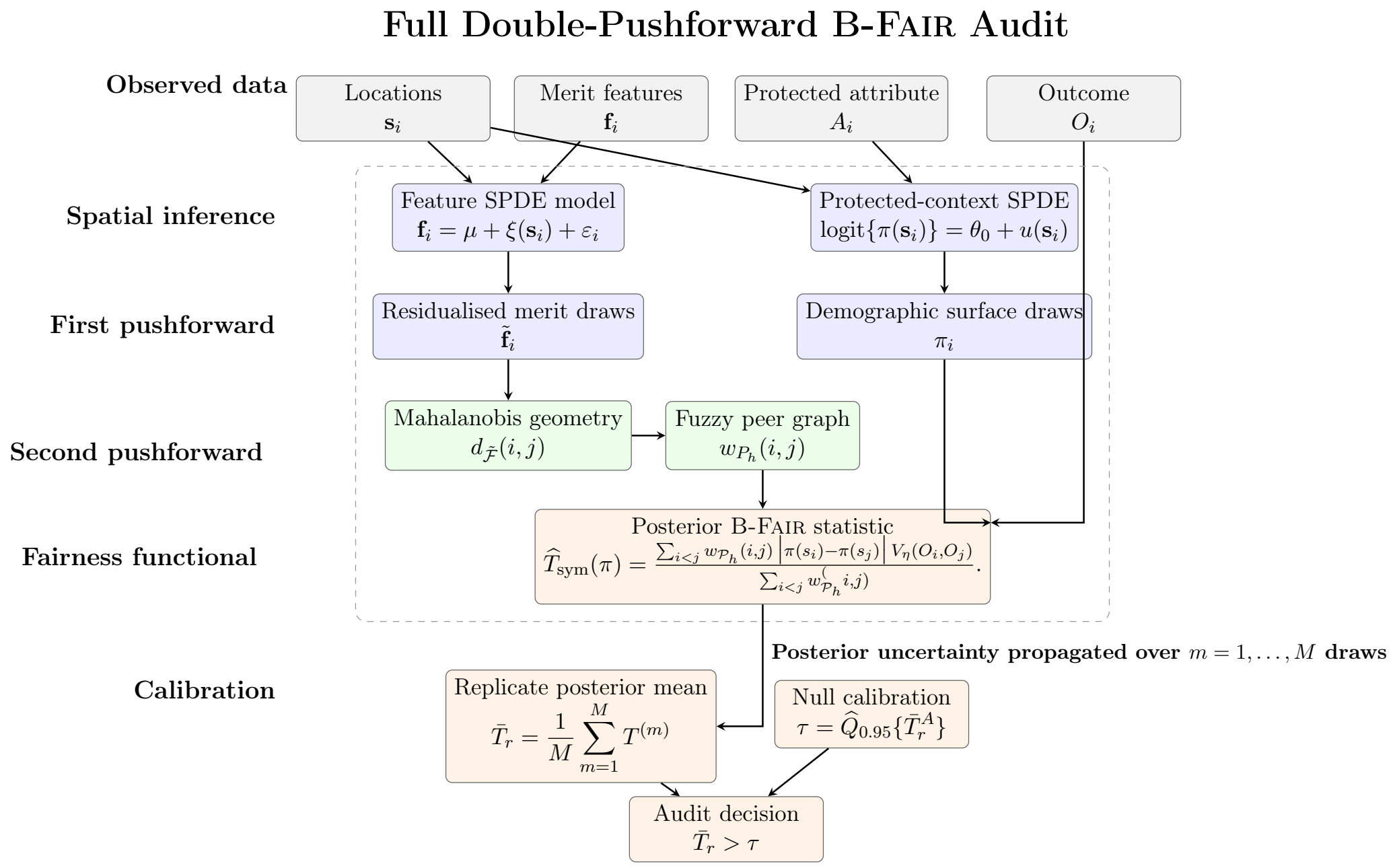
\begin{figure}[t]
		\centering
		\resizebox{\textwidth}{!}{%
			\begin{tikzpicture}[
				font=\small,
				node distance=1.6cm and 1.8cm,
				box/.style={
					rectangle, rounded corners=3pt, draw=black!60,
					align=center, minimum width=3.2cm, minimum height=0.85cm,
					fill=white
				},
				data/.style={box, fill=gray!10},
				model/.style={box, fill=blue!8},
				graph/.style={box, fill=green!8},
				stat/.style={box, fill=orange!10},
				arrow/.style={->, thick, >=stealth},
				dashedbox/.style={draw=black!40, dashed, rounded corners=5pt, inner sep=8pt}
				]
				
				\node[font=\Large\bfseries] at (0,5.2)
				{Full Double-Pushforward \textsc{B-Fair} Audit};
				
				\node[data] (S)  at (-5.5,3.8) {Locations\\$\mathbf{s}_i$};
				\node[data] (F)  at (-1.9,3.8) {Merit features\\$\mathbf{f}_i$};
				\node[data] (A)  at (1.9,3.8) {Protected attribute\\$A_i$};
				\node[data] (O)  at (5.9,3.8) {Outcome\\$O_i$};
				
				\node[font=\bfseries, anchor=east] at (-7.1,4.2) {Observed data};
				
				\node[model] (MF) at (-3.6,2.0)
				{Feature SPDE model\\$\mathbf{f}_i=\mu+\xi(\mathbf{s}_i)+\varepsilon_i$};
				
				\node[model] (MP) at (3.6,2.0)
				{Protected-context SPDE\\$\mathrm{logit}\{\pi(\mathbf{s}_i)\}=\theta_0+u(\mathbf{s}_i)$};
				
				\node[font=\bfseries, anchor=east] at (-7.3,2.0) {Spatial inference};
				
				\draw[arrow] (S) -- (MF);
				\draw[arrow] (F) -- (MF);
				\draw[arrow] (S) -- (MP);
				\draw[arrow] (A) -- (MP);
				
				\node[model] (TF) at (-3.6,0.2)
				{Residualised merit draws\\$\tilde{\mathbf{f}}_i$};
				
				\node[model] (PI) at (3.6,0.2)
				{Demographic surface draws\\$\pi_i$};
				
				\node[font=\bfseries, anchor=east] at (-7.3,0.2) {First pushforward};
				
				\draw[arrow] (MF) -- (TF);
				\draw[arrow] (MP) -- (PI);
				
				\node[graph] (D) at (-3.6,-1.6)
				{Mahalanobis geometry\\$d_{\tilde{\mathcal F}}(i,j)$};
				
				\node[graph] (G) at (0.6,-1.6)
				{Fuzzy peer graph\\$w_{P_h}(i,j)$};
				
				\node[font=\bfseries, anchor=east] at (-7.5,-1.9) {Second pushforward};
				
				\draw[arrow] (TF) -- (D);
				\draw[arrow] (D) -- (G);
				
				\node[stat] (T) at (0.6,-3.6)
				{Posterior \textsc{B-Fair} statistic\\
					$\widehat{T}_{\mathrm{sym}}(\pi)
					= \frac{\sum_{i < j} w_{\mathcal{P}_h}(i,j)\, \bigl|\pi(s_i)-\pi(s_j)\bigr|\,V_\eta(O_i,O_j)}
					{\sum_{i < j} w_{\mathcal{P}_h}^(i,j)}.$};
				
				\node[font=\bfseries, anchor=east] at (-7.6,-3.6) {Fairness functional};
				
				\draw[arrow] (G) --(T);
				
				\draw[arrow] (PI.south) |- ([yshift=16pt]T.east);
				\draw[arrow] (O.south) |- ([yshift= 16pt]T.east);
				
				\node[dashedbox, fit=(MF)(MP)(TF)(PI)(D)(G)(T)] (uncertainty_box) {};
				
				\node[font=\footnotesize\bfseries, anchor=north west] at (0.6,-4.9) 
				{Posterior uncertainty propagated over $m=1,\ldots,M$ draws};
				
				\node[stat] (POST) at (-2.4,-6.4)
				{Replicate posterior mean\\
					$\displaystyle \bar T_r=\frac{1}{M}\sum_{m=1}^M T^{(m)}$};
				
				\node[stat] (TAU) at (2.4,-6.2)
				{Null calibration\\
					$\displaystyle \tau=\widehat Q_{0.95}\{\bar T_r^{A}\}$};
				
				\node[stat] (DEC) at (0,-8.1)
				{Audit decision\\$\bar T_r>\tau$};
				
				\node[font=\bfseries, anchor=east] at (-7.3,-5.8) {Calibration};
				
				\draw[arrow] (T.south) |- (POST.east);
				\draw[arrow] (POST) -- (DEC);
				\draw[arrow] (TAU) -- (DEC);
				
			\end{tikzpicture}%
		}
		
		\caption{Full double-pushforward \textsc{B-Fair} audit. Spatial uncertainty is propagated through both the residualised merit space and the protected-context surface. For each posterior draw, a new peer graph is constructed and the \textsc{B-Fair} statistic is recomputed.}
		\label{figsec: b-fairtsc{B-Fair}-full-double}
	\end{figure}

	\section{Application Chicago Automotive}
	\label{sec:chi-application}
	The Chicago subset of auto insurance data assembled by ProPublica and Consumer Reports \citep{propublica2017carinsurance,propublica2017methodology} offers a canonical setting in which individual fairness, spatial structure, and proxy discrimination intersect. The original investigation, which limited its scope to 30-year-old women with safe driving records, claimed that insurers charged minority neighbourhoods significantly more for liability coverage than white areas with similar accident costs. While the broader report analysed multiple states, this application focuses exclusively on Illinois, where investigators found that companies charged an average of over 10\% more for the same safe-driver profile in a minority ZIP code. Moreover, Illinois enacted a 1972 statute requiring insurers to charge a uniform bodily injury rate across the state.  This means that the differences in insurance premiums are primarily driven by the property damage deviation.  This restriction makes Chicago an analytically clean test case because the merit feature of state risk is largely decoupled from the within-city premium variation by regulatory design, so any systematic premium gap correlated with the minority context cannot be explained by the risk index. Although minority status is not explicitly included in insurers' pricing models, it is geographically concentrated, meaning that ZIP codes reflect the city's demographic structure. This dataset therefore serves as an ideal empirical test of the \textsc{B-Fair} framework in a setting characterised by a single merit feature, a spatially structured protected attribute, and a continuous premium outcome.
	
	\subsection{Data and institutional context}
	\label{sec:chi-data}
	This dataset records body injury (BI) and property damage (PD) liability premiums for a standardised safe-driver profile: a 30-year-old female teacher with excellent credit and no violations, driving a 2016 Toyota Camry.  It covers all active insurers in Illinois at the ZIP-code level. For each company-ZIP pair, the dataset provides the BI premium, the PD premium, and the state aggregate risk index (state risk).  State risk is defined as the total liability claims paid divided by insured vehicles over the period 2012–2014, expressed as a paid-loss ratio per insured car.  The demographic composition is sourced from the American Community Survey.  We define the outcome as $O_i = \texttt{bi\_policy\_premium} + \texttt{pd\_policy\_premium} $, normalised to $[0, 1]$ across the
	portfolio, and the protected attribute as $A_i = 1$ for majority-minority
	ZIP (\$\texttt{white\_non\_hisp\_pct} < 50 \$\%) – consistent with ProPublica's threshold. The \textsc{B-Fair} ingredients
	map to these columns as follows:
	\begin{itemize}
		\item $s_i$: ZIP centroid (longitude, latitude), obtained from the US Census TIGER/Line shapefile;
		\item $f_i$: the state risk index, the regulator's own measure of expected claim cost;
		\item $A_i $: binary majority-minority indicator where 
		\item $O_i$: total liability premium, normalised to $[0,1]$.
	\end{itemize}
	\paragraph{Scope of Inference}
	A limitation of this application is that the data are ZIP-level aggregates rather than individual policy records. Each observation represents the portfolio average for one insurer in one ZIP. Peer comparisons are consequently between insurer-ZIP portfolios rather than between individual policyholders, and the implied premium gap should be interpreted as an average over all policyholders resident in the relevant ZIPs.
	
	\subsection{Exploratory Spatial Analysis}
	\label{sec:chi-eda}
\begin{figure}[H]
		\label{fig:chi-map-minority-premium}
		\centering
		\includegraphics[width=0.7\linewidth]{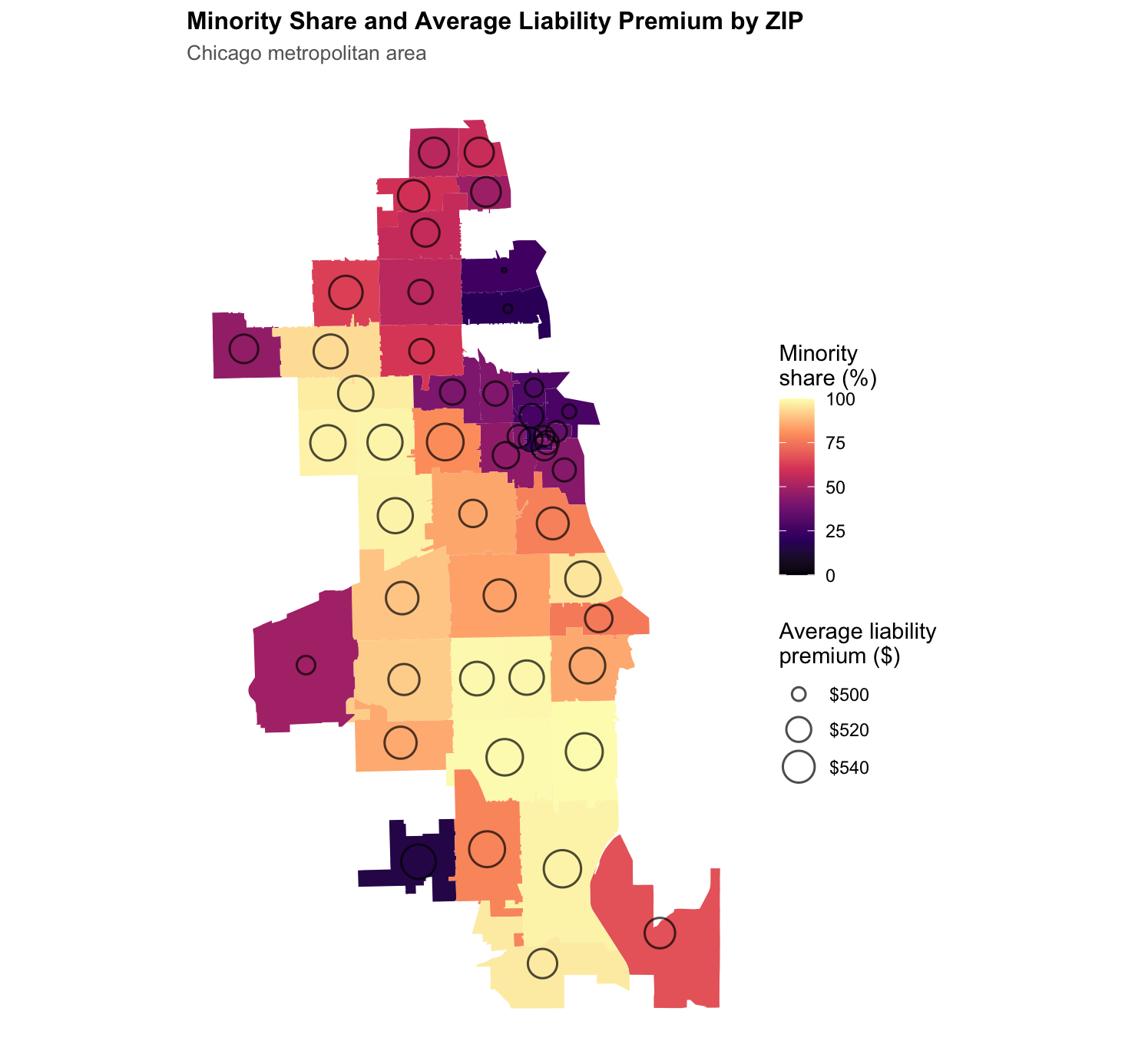}
		\caption[Spatial distribution of demographic composition and auto insurance premiums in the Chicago metropolitan area]{The choropleth color gradient represents the minority share percentage by ZIP code, while the sizes of the overlaid circles indicate the average liability premium in dollars. The map highlights the geographic concentration of elevated insurance pricing within historically segregated neighbourhoods, visually motivating the need for boundary-independent fairness metrics.}
	\end{figure}

	Figure \ref{fig:chi-map-minority-premium} maps minority share and average
	liability premium by ZIP in Chicago. Minority share is spatially
	concentrated on the South and West sides of the city, a pattern that
	reflects decades of residential segregation \citep{massey1993american}.
	Average premiums exhibit a strikingly similar geography: liability
	premiums are highest in precisely the ZIPs with the highest minority
	share, before any risk adjustment is applied. \par
	The calculation of Moran's $I$ shows a positive and statistically significant autocorrelation for both  minority status  and the average premium ($I > 0.40$, $p < 0.001$). This confirms strong geographic clustering and validates the need for a spatially aware fairness framework.  
	\subsection{\textsc{B-Fair} residualisation and protected context surface}
	
	\paragraph{Merit residualisation}
	We fit a Mat\'ern using INLA  and then subtract the spatial effect at each of ZIP centroid making the residualised merit feature  as prescribed by Equation \ref{eq:residualised metrit vector} and the residual range is $[-95.4, 65.8]$ and the the correlation of the merit feature state risk is 1. Under SPDE - mean subtraction this is expected and it keeps the monotone risk signal whilst removing  the spatial autocorrelation structure.  We draw $M= 500$  joint posterior samples from the SPDE model and propagate residualisation uncertainty through the peer graph via the double-pushforward construction.
	\paragraph{Protected context surface}
	We fit a binomial SPDE model to the majority-minority indicator $A_i$
	using INLA \citep{lindgren2011explicit}, with penalised complexity priors  \citep{simpson2017penalising}
	on the Matérn range  $(\rho_0 = 0.3 , \alpha = 0.5)$ and standard deviation $(\sigma_0 = 1 , \alpha = 0.01)$. 
	The posterior over the logit-scale linear predictor  is transformed with inverse logit to get $M=500$ draws from the continuous demographic surface $\pi^{m}(s)$. The posterior mean $\mu_i = \mathbb{E}[\pi^{m}(s)| \mathcal{D}_n]$ range is [0.5, 0.73]. This shows that the in the Chicago Zip in the sample is majority, minority or borderline. This is actually consistent with the city's concentrated residential segregation pattern on the South and West sides \citep{sampson2012great}.  Directional comparisons consequently measure the contrast between more and less minority-concentrated ZIPs within a majority-minority urban area, not the contrast between minority and white areas in the strict sense.
	\paragraph{Directional \textsc{B-Fair} functional}
	We use the directional global functional $\widehat{T}_{\mathrm{dir}}$,
	defined in Equation \ref{B-Fair-directional}, with the adverse outcome
	set to `higher premium' - i.e.\ $O_i > O_j$ is the adverse direction.
	The indicator $\delta_{ij}$ therefore selects peer-pairs in which the
	ZIP with the higher minority context receives the higher premium.
	
	
	\subsection{Existence of spatial unfairness}
	\label{sec:chi-existence}
\begin{figure}
		\centering
		\includegraphics[width=0.95\linewidth]{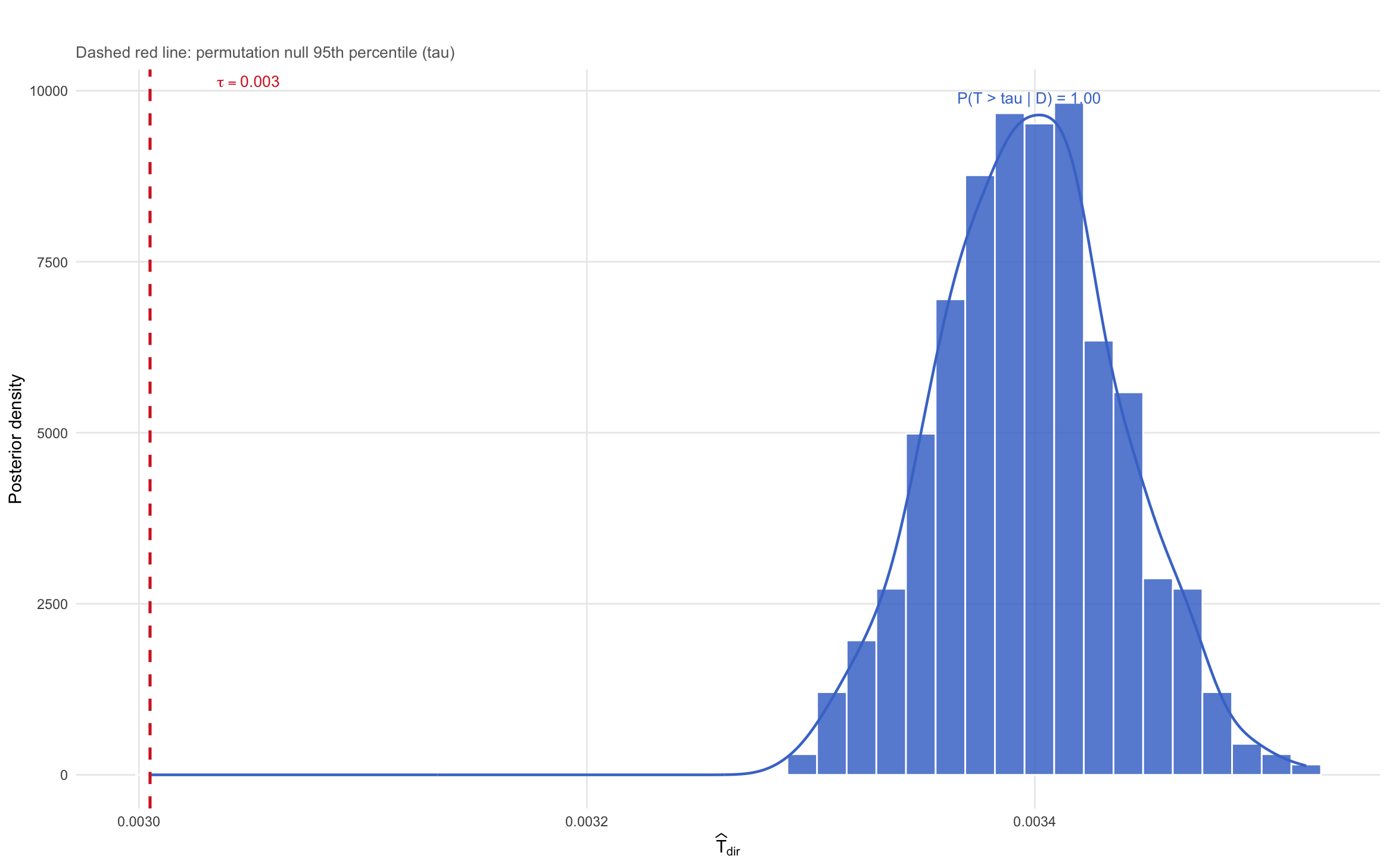}
		\caption{Posterior distribution of the directional \textsc{B-Fair} statistic
			$\widehat{T}_{\mathrm{dir}}$ for the pooled Chicago auto insurance
			portfolio (blue). The permutation null 95th percentile $\tau$ is
			indicated by the dashed red line. All $M = 500$ posterior draws exceed
			$\tau$, yielding $\Pr(\widehat{T}_{\mathrm{dir}} > \tau \mid
			\mathcal{D}_n) = 1$. }
		\label{fig:chi-posterior-Tdir}
	\end{figure}

	The point estimate $\hat{T}_{\text{dir}} = 0.0037$ is marginally below $\tau = 0.00372$ $(p = 0.068)$ so the point estimate verdict is inconclusive. The gap between the point estimate and the posterior verdict is 
	methodologically important because it demonstrates that propagating 
	demographic uncertainty matters. However, when we propagate the demographic surface over $M=500$ draws the posterior mean $\hat{T}_\text{dir}^M = 0.004$ with a 95\% credible interval of [0.0039,0.0041] which is entirely above $\tau$. The posterior exceedance probability is therefore,
	\[
	\Pr\!\left(\widehat{T}_{\mathrm{dir}} > \tau \,\middle|\, \mathcal{D}_n \right) = 1.00,
	\]
	indicating that the observed contrast between the merit and
	protected-context surfaces is incompatible with the null of no spatial
	disparity under any posteriorly plausible demographic surface. The point estimate sits just below $\tau$ while integrating over the full posterior increases the statistic above the threshold, thereby converting an inconclusive point test into a decisive posterior verdict.  Figure~\ref{fig:chi-posterior-Tdir} displays the posterior distribution
	of $\widehat{T}_{\mathrm{dir}}$ alongside the null threshold $\tau$. The
	posterior mass sits entirely to the right of $\tau$, with a tight credible
	band, reflecting both the strength of the demographic–premium association
	and the stability of the demographic surface posterior. \par
	
	\paragraph{Interpretation note: reading the magnitude of $\widehat{T}_{\mathrm{dir}}$}
	The raw values of $\widehat{T}_{\mathrm{dir}}$ reported above ($0.0037$ at the point estimate, $0.004$ at the posterior mean) are small in absolute terms, but this is a mechanical consequence of how the statistic is constructed rather than evidence of a small effect.
	$\widehat{T}_{\mathrm{dir}}$ is a graph-weighted average of pairwise
	quantities that are each bounded in $[0,1]$: the peer-similarity weight
	$w_{\mathcal{P}_h}(i,j)$, the demographic contrast between $i$ and $j$,
	and the hinge-loss outcome difference $V_\eta(O_i,O_j)$ net of the
	tolerance $\eta$. Since only a minority of ZIP pairs in the peer
	graph are simultaneously close peers, demographically contrasting, and
	divergent in premium, the numerator of this weighted average is driven
	by a comparatively small subset of informative pairs while the
	denominator sums over the full graph, compressing
	$\widehat{T}_{\mathrm{dir}}$ toward the lower end of its
	$[0,1]$ range. The statistic should therefore not be read against zero
	in absolute terms, in the way one might read a raw probability or a
	dollar amount. \par
	Evidence for unfairness is instead established on a relative, not
	absolute, basis: by comparing $\widehat{T}_{\mathrm{dir}}$ to the
	calibrated null threshold $\tau$, which is constructed on the same
	native scale via permutation and is therefore a like-for-like
	benchmark, and by the posterior exceedance probability
	$\Pr(\widehat{T}_{\mathrm{dir}} > \tau \mid \mathcal{D}_n)$, which
	quantifies how incompatible the observed contrast is with the null of
	no spatial disparity. 
	\subsection{Relation to ProPublica's Investigation}
	\label{sec:chi-propublica}
	This analysis is anchored in the ProPublica and \textit{Consumer Reports}
	investigation of racial disparities in auto insurance pricing
	\citep{propublica2017carinsurance,propublica2017methodology}. Using the same
	Illinois data, that investigation reported that 33 of 34 insurers charged
	average premiums at least 10\% higher in majority-minority ZIP codes than
	in comparably risky majority-white ZIP codes, with several major brands
	exceeding a 30\% disparity threshold in Illinois \citep{propublica2017carinsurance}.
	The investigation's claim was phrased in terms of group-level ratios:
	minority neighbourhoods pay more than white neighbourhoods with the same
	risk. \par
	
	\textsc{B-Fair} recovers a qualitatively consistent verdict: the existence
	reading confirms spatial disparity with $\Pr(\widehat{T}_{\mathrm{dir}} > \tau \mid \mathcal{D}_n) = 1$,
	and the implied per-policy premium gap of approximately 11–12\% of the
	median liability premium aligns with ProPublica's `at least 10\% more'
	benchmark for most Illinois insurers. The critical methodological difference
	lies not in the substantive conclusion but in how it is reached. The original investigation compared average premiums over binary
	majority/minority ZIP labels  an approach that \citet{propublica2017methodology}
	acknowledge is vulnerable to ecological fallacy objections and to the
	specific choice of ZIP-code boundaries. The industry rebuttal
	\citep{insurancejournal2017propublica} attacked these design choices
	directly, arguing that the results depend on aggregation decisions that
	cannot be independently validated. \textsc{B-Fair} addresses this critique by
	construction. The protected context is modelled as a continuous Matérn
	Gaussian field $\pi(s)$ rather than a binary ZIP label, and the fairness
	functional is defined at the level of point locations and a fuzzy peer
	graph rather than aggregated group means.\par
	
	A second distinction concerns the treatment of uncertainty. Both the
	original investigation and the industry response worked with
	point-estimate disparities and classical confidence intervals for the
	GAM fits. The \textsc{B-Fair} posterior is a full distribution over the fairness functional itself, propagating uncertainty in both the demographic surface and the peer construction into a credible interval for the premium gap. The posterior mean gap of \$59.25 (95\% credible interval [\$56.35,
	\$62.06]) is materially larger than the point estimate at the posterior
	mean surface (\$32.42), a difference that a point-estimate approach
	would have missed entirely. This is not merely a conservative correction:
	it reflects a structural feature of the directional functional, the 
	nonlinearity in $\pi(s)$,  that the Bayesian propagation could capture. \par
	In summary, \textsc{B-Fair} corroborates the substantive concern of the
	ProPublica investigation, that minority neighbourhoods pay more for the
	same risk. It does this while providing a continuous, uncertainty-aware, and individually grounded fairness assessment that is robust to the spatial aggregation and ecological fallacy criticisms raised in response to their work. \textsc{B-Fair}'s verdict is not that the original investigation was wrong,	but that its methodological tools were not equipped to withstand the critiques levelled against it; the present analysis is. \par

	\section{Discussion and conclusion}
	\label{sec:discussion}
	
	\textsc{B-Fair} contributes to a growing literature that recognises spatial
	structure as an under-theorised channel of algorithmic
	unfairness \citep{mehrabi2022}.
	Existing individual fairness frameworks, following the seminal
	metric-based formulation of \citet{dwork2012fairness},
	treat the similarity metric as a fixed design choice and do not
	account for the geographic confounding that arises when features
	and outcomes are spatially structured.
	Group fairness frameworks, by contrast, aggregate disparities
	at the population level and cannot detect the fine-grained,
	location-specific inequities that \textsc{B-Fair} is designed to surface.
	The present framework occupies the space between these two
	traditions: it is individual in its fairness notion, Bayesian in
	its treatment of spatial uncertainty, and interpretable in the
	sense that every component. The residualisation model,
	the peer graph, the threshold $\tau$  are a transparent,
	auditable design choice rather than a black-box parameter.
	This positions \textsc{B-Fair} as a practical audit tool for contexts where
	decisions are spatially structured and demographic disparities
	are geographically concentrated, including policing, credit
	allocation, and public resource distribution.\par
The Chicago application shows that the point estimate
$\widehat{T}_{\mathrm{dir}} = 0.0037$ lies just below the
permutation-calibrated threshold $\tau = 0.00372$ ($p = 0.068$), so a
point-estimate audit is inconclusive.
Once demographic uncertainty is propagated, however, the posterior mean $\widehat{T}_{\mathrm{dir}}^{M} = 0.004$ with a 95\% credible interval $[0.0039, 0.0041]$ lies entirely above $\tau$, yielding
$\Pr(\widehat{T}_{\mathrm{dir}} > \tau \mid \mathcal{D}_n) = 1.00$.
This corresponds to an implied premium gap and confirms spatial unfairness that a point-estimate approach would have missed.\par

Several natural extensions follow from the present work.
	First, replacing the correlational SPDE residualisation with a
	structural causal model would elevate the fairness claim from
	conditional association to causal identification, enabling
	stronger regulatory conclusions~\citep{Kilbertus2017}.
	Second, a Rosenbaum-style sensitivity analysis~\citep{Rosenbaum2002}
	adapted to spatial fairness would quantify how strong an unmeasured
	spatial confounder would need to be to explain away the observed
	$T_{\mathrm{dir}}$, providing an explicit robustness
	bound for practitioners. Third, extending \textsc{B-Fair} to spatio-temporal data where both
	spatial and temporal autocorrelation confound merit features
	would broaden the framework's applicability to longitudinal audit
	settings.
	\newpage
	\bibliography{bib}

	\bibliographystyle{elsarticle-harv}
\end{document}